\documentclass{article}
\usepackage{graphicx}

\newcommand{\tr}{\ensuremath{\mathop{\mathrm{tr}}\nolimits}}
\newcommand{\mn}{{\mu\nu}}

\newcommand{\duF}{\tilde{F}}

\newcommand{\htt}{\mathbf{T}}
\newcommand{\be}{\begin{equation}} \newcommand{\ee}{\end{equation}}

	\title{
\begin{flushright}\begin{small}
	DTP-MSU/03-03\\
	hep-th/0301069	
\end{small}\end{flushright}	
Stabilization of the Yang-Mills  chaos in non-Abelian Born-Infeld theory} 

\author{ D.V. Gal'tsov\thanks{galtsov@grg.phys.msu.su} ~and
 V.V. Dyadichev\thanks{vlad@grg1.phys.msu.su}\\\\
\textit{Department of Theoretical Physics,}\\
\textit{Moscow State University, Moscow, Russia} } 

\date{\today}
\begin{document}

\maketitle 
\begin{abstract}
	We investigate dynamics of the homogeneous
time-dependent SU(2) Yang-Mills fields governed by the non-Abelian
Born-Infeld lagrangian which arises in superstring theory as a
result of summation of all orders in the string slope parameter
$\alpha'$. It is shown that generically the Born-Infeld dynamics
is less chaotic than that in the ordinary Yang-Mills theory, and
at high enough field strength the Yang-Mills chaos is stabilized.
More generally, a smothering effect of the string non-locality on
behavior of classical fields is conjectured.
\end{abstract}

\maketitle

The chaotic nature of the classical Yang-Mills dynamics was
recognized more than twenty years ago
\cite{BaMaSa79,ChSh81,ChSh82,Bibook,Ma00}. It can be revealed
already on simple models of the homogeneous fields depending only
on time \cite{MaSaTe81}. Further studies have shown persistence
of chaotic features in classical dynamics of spatially varying
Yang-Mills fields   leading  to a new kind of the ``ultraviolet
catastrophe'' \cite{NiRuRu96}.

Here we would like to investigate whether this tendency is
preserved on a deeper level of superstring theory, or the latter
provides a smothering effect. The possibility to explore this
question quantitatively is related to  existence of a closed form
effective action for the open strings or the stuck of D-branes
which accumulates all orders in the string slope parameter
$\alpha'$. Strictly speaking, an exact in $\alpha'$ effective
action is known only for the U(1) gauge field, in which case it
is the Born-Infeld (BI) action \cite{FrTs85,Le89}. In the
non-Abelian case the exact calculation of the effective action is
not possible, but a certain non-Abelian version of the Born-Infeld
(NBI) action can still be envisioned in a closed form. This
problem was extensively discussed recently
\cite{Ts97,Ts99,Gonorazky:1998ay}, one reasonable proposal
(though perhaps still not giving the precise answer in the sense
of superstring theory)   involves a symmetrized trace \cite{Ts97}
construction of the action. Another, simpler form, uses a direct
non-Abelian generalization of the U(1) BI action, we will call it
the ordinary trace model. Investigation of static solitons in NBI
theory \cite{Dyadichev:2000iw} has shown qualitative agreement
between the results of both the symmetrized trace and the
ordinary trace models, so here, for the reason of simplicity, we
adopt the latter one:
\begin{equation}
  \label{eq:trbi}
  S_{NBI}=\beta^2\int\,d^4 x\left(1-\sqrt{1+\frac 1{2\beta^2}
  F^a_\mn F_a^\mn-\frac{1}{16\beta^4}(F^a_\mn
  \duF_a^\mn)^2}\right),
\end{equation}
 with $\beta$ being the critical BI field, in string theory
 $\beta=1/2\pi\alpha'$.

The simplest non-Abelian configuration for which the ordinary YM
theory predicts chaotic behavior  \cite{MaSaTe81} is the following
\begin{equation}
A=u\htt_1 dx + v \htt_2 dy, \label{savans}
\end{equation}
where $u$ and $v$ are functions of time only, and
$\htt_1$,$\htt_2$ are the gauge group generators. The
corresponding field strength
\[ F=\dot{u} \htt_1
dt\wedge dx + \dot{v} \htt_2 dt \wedge dy + uv\htt_3 dx \wedge dy,
\]
contains both electric and magnetic components, but these are
mutually orthogonal, so the pseudoscalar invariant  $\tr F\duF$,
 generically  dominant (\ref{eq:trbi}) at high field
strength, is equal to zero. The corresponding one-dimensional
lagrangian reads \be
L=\beta^2\left(1-\sqrt{1-\beta^{-2}\left(\dot{u}^2 +
    \dot{v}^2-v^2u^2\right)}\right) \label{3dnbi}.  \ee

 The equations of motion have the form
\begin{eqnarray}
\ddot{u}&=&-u v^2+\frac{2 \dot{u}vu(\dot{u}v+\dot{v}u ) }{\beta^2+u^2v^2},
\label{4}\\
\ddot{v}&=&-v u^2+\frac{2 \dot{v}vu (\dot{u}v+\dot{v}u )
}{\beta^2+u^2v^2}.\label{5}
\end{eqnarray}
In the limit $\beta\to\infty$ of the ordinary YM action  the
second terms in the right hand sides of these equations vanish
and one obtains the ``hyperbolic
billiard''\cite{MaSaTe81,AsSa83,DaRu90}, known in various
problems of mathematical physics, which is a two-dimensional
mechanical system with the potential $u^2 v^2$. This potential has
two valleys along the lines $u=0$ и $v=0$, the particle motion
being confined by hyperbolae  $uv= \mathrm{const}$. The existence
of these valleys is crucial for emergency of chaos.

The trajectories of the   system governed by the BI action depend
on the energy integral   $E$ and the parameter $\beta$. Rescaling
of the field variables together with time rescaling maps
configurations with different $E$ and $\beta$ onto each other, so
it will be enough  to consider dependence of the dynamics only on
$\beta$ assuming for the energy any fixed value, e.g. $E=1$. One
simple (though non-rigorous) method to reveal chaoticity is the
analysis of the geodesic deviation equation for a  suitably
defined pseudoriemannian space whose geodesics coincide with the
trajectories of the system \cite{Sz93,SoSuMa96}. The lagrangian
(\ref{3dnbi}) gives rise to the following metric tensor \be
ds^2=\left(1+\frac{u^2v^2}{\beta^2}\right)dt^2-
\frac{1}{\beta^2}(du^2+dv^2), \label{effmetr} \ee which is
regular for all finite $\beta\neq 0$.  Consider the deviation
equation for two close geodesics   \be \frac{D^2n^i}{ds}=
R^i{}_{jkl}u^ju^kn^l, \ee where $D^2n^i$ is the covariant
differential of their transversal separation, $R^i{}_{jkl}$ is the
Riemann-Christoffel tensor, and
$u^j$ is the three-speed in the space (\ref{effmetr}). Locally,
the geodesic deviation is described by the matrix
$R^i{}_{jkl}u^ju^k$ depending on points $u,v,\dot{u},\dot{v}$  of
the phase space. If at least one eigenvalue of this matrix is
positive, the geodesics will exponentially diverge with time.
Negative eigenvalues correspond to oscillations or to slower
divergence, e.g. power-law.  In our case one of the eigenvalues
of the matrix  $R^i{}_{jkl}u^ju^k$ vanishes (due to the static
nature of the metric), two others are the roots of a quadratic
equation and read
\begin{equation}\label{lam}
 \lambda_{1,2}=\frac12\left(\mathcal{B}
 \pm\sqrt{\mathcal{B}^2+4\mathcal{C}}\right),
\end{equation}
where
\begin{eqnarray}
\mathcal{B}&=&\frac{2u v\dot{u}\dot{v}-\beta^2(u^2+v^2)}{\beta^2
U W}+
\frac{(\dot{v}u+\dot{u}v)^2}{\beta^2 U^2 W},\\
\mathcal{C}&=&\frac{v^2 u^2(3\beta^2+u^2 v^2)}{\beta^2 U^2 W },
\end{eqnarray}
with   $W$ and   $U$ being  positive functions
\[ W= 1+\frac{u^2v^2-\dot{u}^2-\dot{v}^2}{\beta^2},\qquad
U=1+\frac{u^2v^2}{\beta^2}.\] It is easy to see that both
non-zero eigenvalues (\ref{lam}) are real, one  positive and the
other negative. Therefore, for any finite $\beta$, there exist
locally divergent geodesics, and it can be expected that for any
$\beta$ the motion will remain chaotic. But in the limit
$\beta\to 0$ these eigenvalues tend to zero and the analysis
becomes inconclusive. Qualitatively, this phenomenon of decrease
of the geodesic divergence with decreasing $\beta$ can be
attributed to the lowering of the degree of chaoticity. It is
also worth noting that for $\beta \to 0$ the second terms in the
equations of motion (\ref{4}-\ref{5}) look like singular friction
terms.

Another simple tool in the analysis of chaos is the calculation
of the Lyapunov exponents defined as
\begin{equation}
  \chi=\lim_{t\to\infty} \frac{1}{t} \log \frac{|\delta x(t)|}{|\delta x(0)|},
\end{equation} where $\delta x(t)$~is a solution of the linearized
perturbation equation along the chosen trajectory. Here $x$
stands for the (four-dimensional) phase space coordinates of the
original system and the Cartesian metric norm is chosen. The
positive value of $\chi$ signals that the close trajectories
diverge exponentially with time so the motion is unstable.  Stable
regular motion corresponds to zero Lyapunov exponents.

The third standard tool is the construction of Poincare sections.
Recall that the Poincare section is some hypersurface in the
phase space.  The phase space for the system (\ref{axisnbi}) is
four-dimensional, but the energy conservation restricts the
motion to a three-dimensional manifold. Imposing one additional
constraint will fix a two-dimensional surface, and we can find a
set of points corresponding to intersection of the chosen
trajectory with this surface. For the regular motion one
typically gets some smooth curves, while in the chaotic case the
intersection points fill finite regions on the surface.

Both methods reveal that with decreasing $\beta$ the degree of
chaoticity is diminished, though we were not able to perform
calculations for very small $\beta$ to decide definitely whether
the chaos is stabilized or not. But in fact the ansatz
(\ref{savans})  is not generic enough from the point of view of
the BI dynamics, since the leading at high field strength term
$(F\tilde F)^2$ in (\ref{eq:trbi}) is zero for it. So we can hope
to draw more definitive conclusions by exploring another ansatz
for which the invariant $(F\tilde F)$ is non-zero. Such an
example is provided by  a simple axially symmetric configuration
of the $SU(2)$ gauge field which is also parameterized by two
functions of time
  $u,\,v$ \cite{DaKu96}:
\begin{equation} A=\htt_1 u dx +\htt_2
  u dy + \htt_3 v dz \label{kunans}.
\end{equation}
The field strength contains both electric and magnetic components
\[
F=\dot{u}(\htt_1 dt\wedge dx +\htt_2 dt\wedge dy) +\dot{v} \htt_3
dt\wedge dz + u^2\htt_3 dx\wedge dy + uv (\htt_1 dy\wedge
dz+\htt_2 dz\wedge dx),
\]
and now the pseudoscalar term $(F\tilde F)^2$  is non-zero.
Substituting this ansatz into the action (\ref{eq:trbi}) we
obtain the following one-dimensional lagrangian:
\begin{equation}\label{axisnbi}
 L_1=\beta ^2   (1- \mathcal{R}),\quad \mathcal{R}=\sqrt{1 -
  \frac{2\dot{u}^2+\dot{v}^2-u^2(u^2+2v^2)}{\beta^2}
 -\frac{u^2(2\dot{u}v+\dot{v}{u})^2}{\beta^4}}.
\end{equation}
In the limit of the usual Yang-Mills theory $\beta\to \infty $ we
recover the system which was considered in \cite{DaKu96}:
\begin{equation}
\label{axiym} L_{YM}=\frac12 (2\dot{u}^2+\dot{v}^2-u^4-2 u^2 v^2).
\end{equation}
The corresponding potential has valleys along $v=0$. Like in the
case of the hyperbolic billiard with potential $v^2u^2$, the
dynamics governed by the usual quadratic Yang-Mills lagrangian
motion exhibits chaotic  character.

However, for the NBI action with finite   $\beta$ the situation is
different. One can use again the geodesic deviation method, the
corresponding  metric being:
\begin{equation}
ds^2=\left(1+\frac{u^4+2u^2v^2}{\beta^2}\right)dt^2-\frac{du^2+2
dv^2}{\beta^2}-\frac{u^2(2v du + u dv)^2}{\beta^4}.
\end{equation}
The energy integral is given by the expression\begin{equation}
E=\frac{\beta^2+u^4+2v^2u^2}{ \mathcal{R}}-\beta^2.
\end{equation}
One can  calculate the eigenvalues of the geodesic deviation
operator as before. In general one finds that the larger
eigenvalue is smaller than one in the case of the previous ansatz
(\ref{savans}). Now the regions in the configuration space appear
in which both non-zero eigenvalues are negative, this corresponds
to locally stable motion. With   decreasing $\beta$ the relative
volume of the local stability regions increases,
 though for every $\beta$  there still exist the regions of local
instability as well. To obtain more definitive conclusions we
performed numerical experiments using Lyapunov exponents and
Poincare sections. The results look as follows.
 For large $\beta$ the dynamics of the system
(\ref{axisnbi}) is qualitatively similar to that in the ordinary
Yang-Mills theory  (\ref{axiym}) and exhibits typical chaotic
features. The trajectories  enter deep into the valleys along
$u=0$. The Poincare sections consist of the clouds of points
filling the finite regions of the plane, while the  Lyapunov
exponents   remain essentially positive. The situation changes
drastically with decreasing $\beta$. For some value of this
parameter, depending on a particular choice of the initial
conditions, one observes  after a series of bifurcations  a clear
transition to the regular motion. The points on the Poincare
sections  $u=0$ line up along the smooth curves (sections of a
torus), while the Lyapunov exponent goes to zero. The motion
becomes quasiperiodic.  A typical feature of such regular motion
is that the trajectory no more enters the potential valleys.  Some
Poincare sections illustrating this chaos-order transition   are
presented on Figs.~1--3. The Lyapunov exponents corresponding to
the first and the last picture of this series are shown on Figs.
4 and  5 respectively.

Comparing with our first system (\ref{savans}) we can guess that
the substantial role in the chaos-order transition  in the NBI
theory is played by the pseudoscalar term in the action
(\ref{eq:trbi}). Since generically this term does not vanish, it
is tempting to say that the transition observed is a generic
phenomenon in the non-Abelian Born-Infeld theory. The effect is
essentially non-perturbative in $\alpha'$, and we conjecture that
it reflects the typical smothering effect of the string
non-locality on the usual stiff field-theoretical behavior. A
related question is whether the cosmological singularity, which
was recently shown to be chaotic at the supergravity level of
string models \cite{DaHe}, will be also regularized when higher
$\alpha'$ corrections are included. Unfortunately there is no
closed form effective action for the closed strings analogous to
the Born-Infeld action for the open strings. But combining
gravity in the lowest order in $\alpha'$ with an exact in
$\alpha'$ matter action we can probe the nature of the
cosmological singularity too. This will be reported in a separate
publication.

The work was supported by RFBR grant 00-02-16306.


\newpage

\begin{figure}[p]
\begin{minipage}[t]{0.48\textwidth}
  \includegraphics[width=\textwidth]{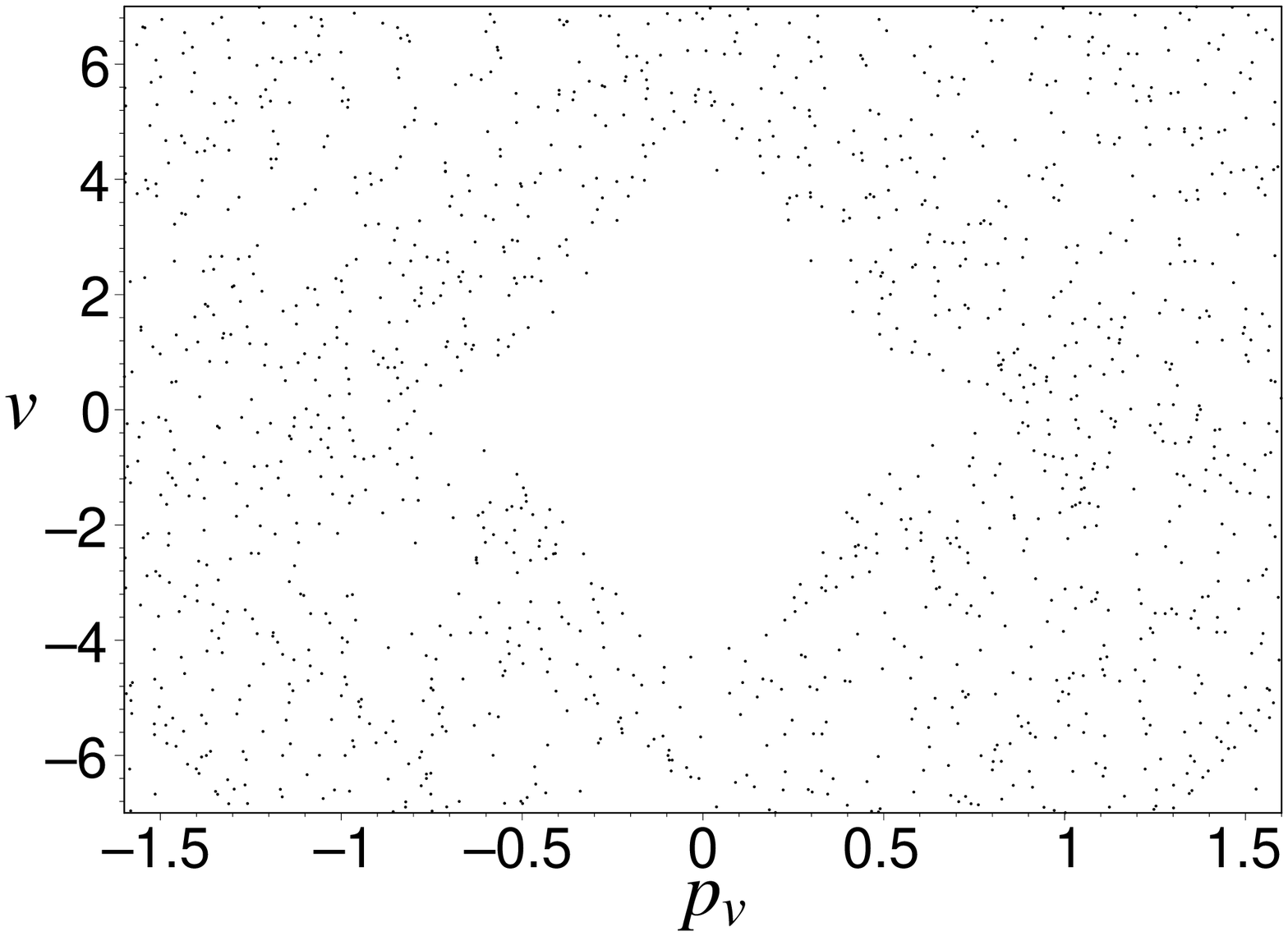}
\caption{\label{fig:poinc3}
Poincare section of the system (\ref{axisnbi}) for
$\beta=0.32$}
\end{minipage}
\begin{minipage}[t]{0.48\textwidth}
  \includegraphics[width=\textwidth]{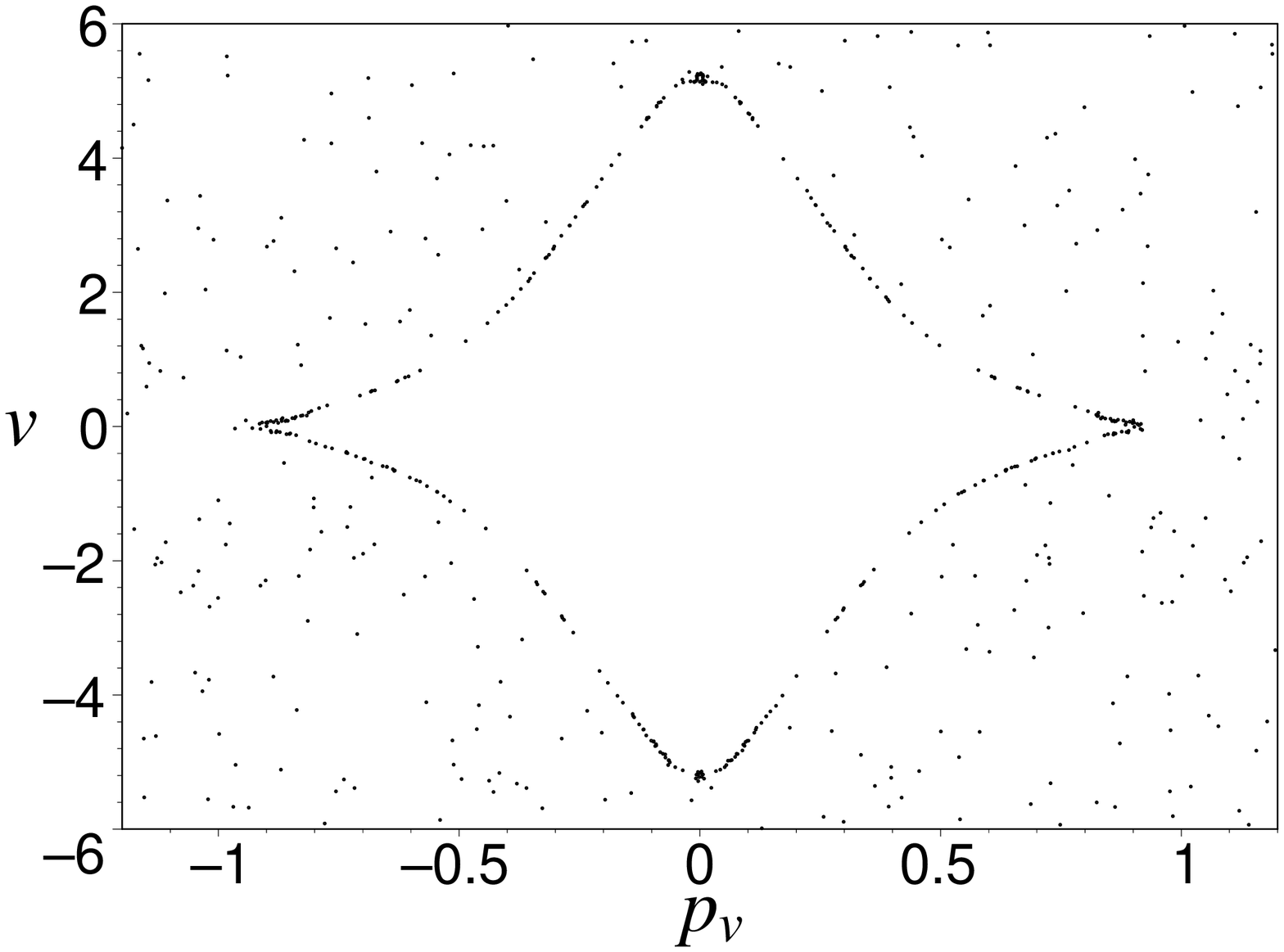}
\caption{\label{fig:poinc2}
$\beta=0.317$}
\end{minipage}
\end{figure}

\begin{figure}[p]
\begin{minipage}[t]{0.48\textwidth}
  \includegraphics[width=\textwidth]{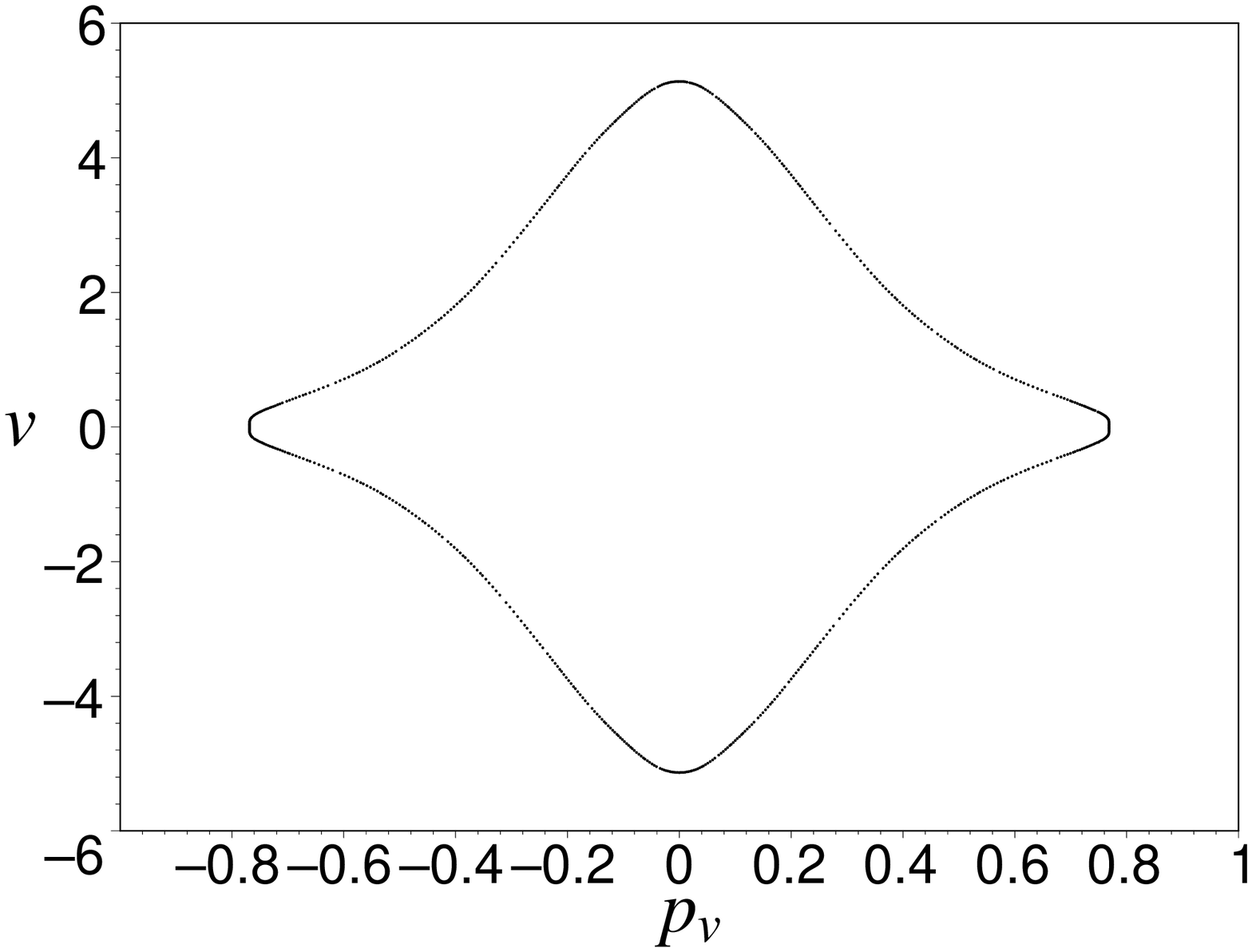}
\caption{\label{fig:poinc1}
$\beta=0.31$}
\end{minipage}
\begin{minipage}[t]{0.48\textwidth}
  \includegraphics[width=\textwidth]{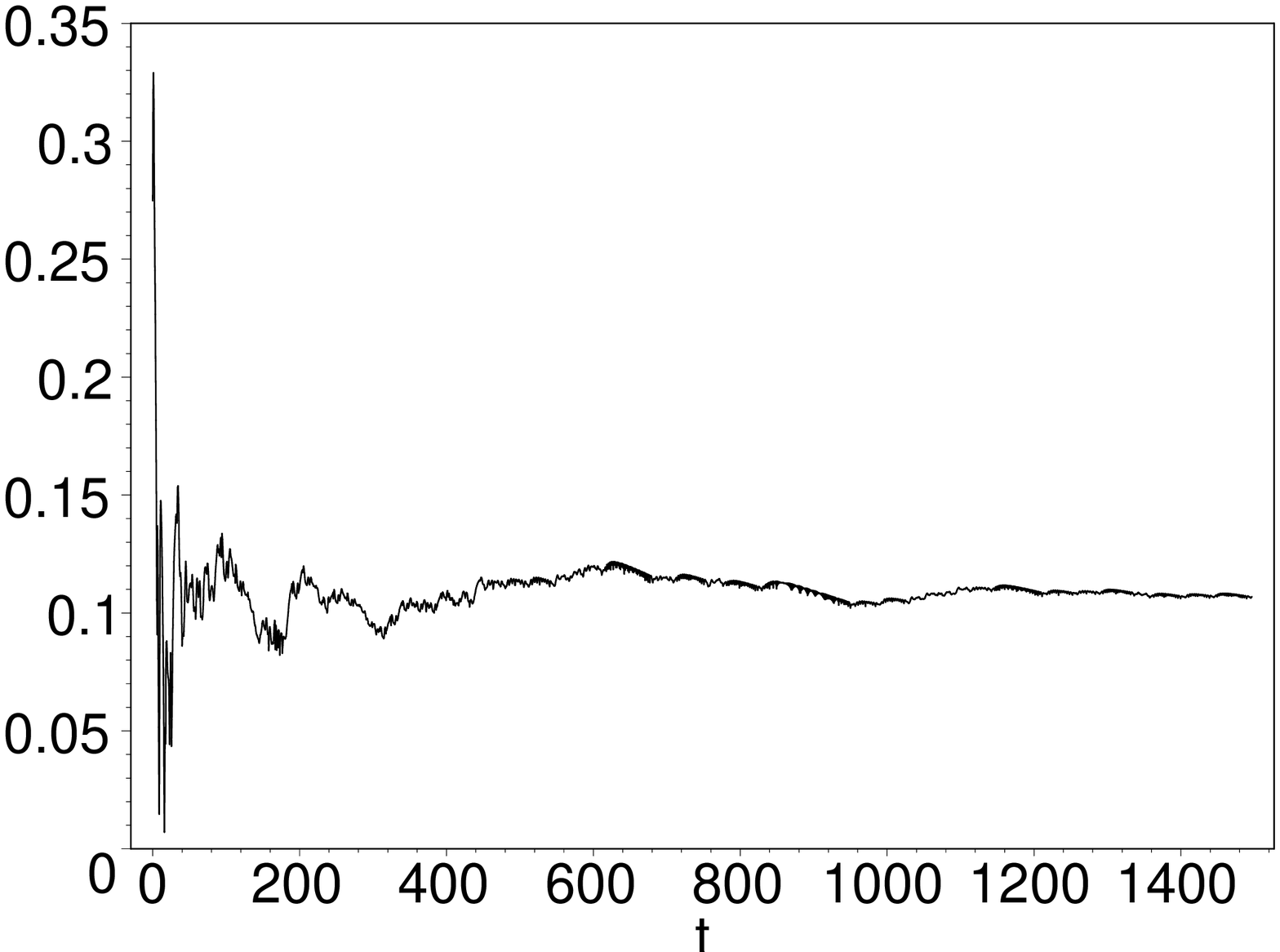}\label{fig:lyap2}
\caption{Lyapunov's exponent corresponding to Fig.
\ref{fig:poinc3}}
\end{minipage}
\end{figure}
\begin{figure}[p]
\begin{minipage}[t]{0.48\textwidth}
  \includegraphics[width=\textwidth]{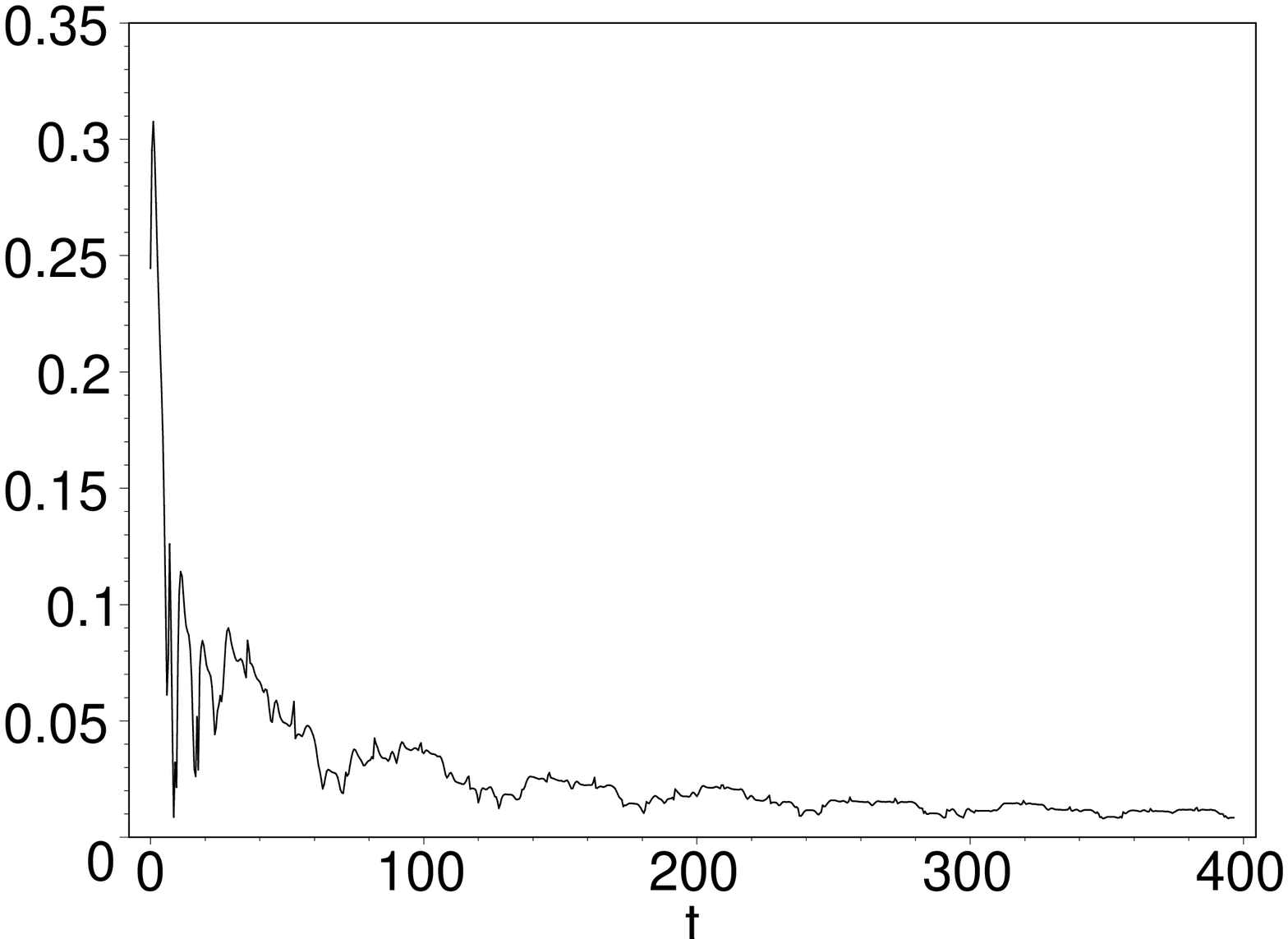}
\caption{Lyapunov's exponent corresponding to Fig.
\ref{fig:poinc1}
\label{fig:lyap1}}
\end{minipage}
\end{figure}

\end{document}